# First evidence for a $Sm^{3+}$-type contribution to the magnetic form factor in the quasielastic spectral response of intermediate valence $SmB_6$

P. A. Alekseev[1,2)], J.-M. Mignot[3)], P.S. Savchenkov[2,1)], V.N. Lazukov[1)]

1) National Research Centre "Kurchatov Institute", 123182 Moscow, Russia
E-mail: pavel_alekseev-r@mail.ru
2) National Research Nuclear University MEPhI, 115409 Moscow, Russia
3) Laboratoire Léon Brillouin - UMR12 CNRS-CEA, CEA Saclay, 91191 Gif-sur-Yvette, France

**Abstract**

The momentum transfer dependence of the magnetic form factor associated with the quasielastic spectral component in the dynamic magnetic response of intermediate valence $SmB_6$ has been measured using inelastic neutron scattering on a double-isotope ($^{154}Sm$, $^{11}B$) single crystal. The experimental dependence differs qualitatively from those obtained earlier for the inelastic signals, as well as from the field-induced magnetic form factor of $SmB_6$ obtained by polarized neutron diffraction. This observation is interpreted by specifically considering the Curie-type contributions to the dynamic susceptibility, which arise from the mixing of $4f^5$ and $4f^6$ $J$-multiplets into the intermediate valence state wave function.

## 1. Introduction

One of the main problems for understanding the magnetic properties of intermediate valence (IV) rare-earth (RE) compounds is to determine the microscopic form of their *f*-electron wave function in the quantum mechanically admixed ground state and the nature of the low-lying excitations. Neutron diffraction and spectroscopy techniques have been used extensively in the study of such materials [1,2,3] because they provide direct insight into both the static and dynamic properties of the magnetic 4*f* shell. In particular, neutron magnetic cross sections can be expressed in terms of the so-called magnetic form factor (MFF), containing information on the dependence of the magnetic neutron scattering intensity on the momentum transfer $Q$ [2,3]. This dependence results from interference effects [4,2], which are sizable because the thermal neutron wavelength is comparable with the radius of the electron shell, of the order of 0.1nm. For localized *f*-electron systems, the MFF can be represented, to a good approximation [4], by a scalar function $f(Q)$ equal to the Fourier transform of the radial distribution of the magnetization density.

An IV state can be regarded as the mixing of two ionic configurations having a different number of electrons, respectively $n$ and $n$-1, on their inner 4*f* shells. It is also thought to involve a partial delocalization of the associated *f*-electron orbitals due to their hybridization with conduction electron states. Both effects can, in principle, be reflected into the $Q$ dependence of the MFF. Its experimental determination was therefore a primary focus of early neutron studies on IV materials.

However, measurements performed on the two archetype IV compounds [5] $SmB_6$ and "collapsed" SmS (obtained by Y doping or application of pressure above 0.6 GPa) led to quite unexpected results [6,7]. Their MFF, obtained by polarized neutron diffraction ("flipping ratios") in an applied magnetic field, failed to show any evidence of a $Sm^{3+}$ contribution, even though the average valence of those materials is known to be close to 2.5. According to both calculations and experiments [2,8], the MFF of $Sm^{3+}$ stands out amongst those of other lanthanide ions, including the divalent species of Sm itself. Its $Q$ dependence is characterized by a maximum around ~ 5 Å$^{-1}$, contrasting with the monotonic decrease found in all other cases. The authors of Ref. [7] tentatively interpreted their observation by assuming temperature dependent crystal electric field (CEF) effects in Sm$^{3+}$ occurring together with strong spin fluctuations, though these two mechanisms are normally thought to be mutually exclusive. Furthermore, the extremely low value (about 0.02$\mu_B$) of the field-induced Sm magnetic moment derived from the polarized neutron diffraction measurement on SmB$_6$ [6] practically rules out a sizable contribution from an ionic Sm$^{3+}$ state, for any hypothetical CEF scheme whatsoever.

The above two systems have also been systematically studied by inelastic neutron scattering (INS). The spin-orbit transitions [$J =0 \to J=1$] of Sm$^{2+}$ and [$J=5/2 \to J=7/2$] of Sm$^{3+}$ have been observed in the form of broad excitation peaks at energies $E$ of 36 meV and 125 meV, respectively [9-14]. Their linewidths reflect spin fluctuations with characteristic time scales on the order of 10$^{-12}$–10$^{-13}$ seconds.

In SmB$_6$, the magnetic response at low temperature ($T \ll 50$ K) further contains a very remarkable structure, consisting of a spin gap region followed by a sharp resonance-like peak at 15 meV [15,9-13]. Upon increasing temperature to about 80 K, the spin gap is gradually suppressed and a quasielastic (QE) signal is recovered. This unusual behavior has been ascribed to the formation of a new bound state corresponding to the quantum mechanical wave function in the IV regime [11,16].

In this work, we examine the different contributions to the SmB$_6$ spectral response, from the standpoint of their $Q$ dependences, as derived from elastic and INS experiments. We discuss possible reasons for the different behaviors observed and, in particular, we report the first observation of a Sm$^{3+}$-like contribution to the MFF associated with the (QE) component in the magnetic spectrum measured at $T = 100$ K. This component had not been studied so far for either SmS or SmB$_6$. Because it is representative of the Curie term in the dynamical susceptibility, it is directly related to the spin-relaxation (fluctuations) due to interaction of the $f$-electrons with their surroundings, a key issue in the physics of IV systems.

The data presented hereafter have been obtained on the same double-isotopic single crystal of $^{154}$Sm$^{11}$B$_6$ used previously in the measurements of Refs. [6,11,12]. The measurements were performed on the thermal neutron three-axis spectrometer 2T at the LLB-Orphée in Saclay.

### 2.1 MFF in neutron scattering measurements

Before presenting our measurements, we recall some important results concerning MFFs in lanthanide compounds. The reader is referred to the literature, in particular the review articles by Lander [2], Holland-Moritz and Lander [3], and Osborn *et al.* [17] for a more comprehensive discussion. RE elements are characterized by a large orbital contribution, and their spin-orbit coupling is typically one order of magnitude stronger than CEF interactions [18]. Russel-Saunders "$L$-$S$" coupling scheme applies and one can use a $|\alpha SLJm\rangle$ representation basis for the electron wave functions. The magnetization density consists of two components, spin and orbital. For convenience, we limit ourselves to the dipole approximation, corresponding to a free ion with spherical symmetry, which is strictly valid at small $Q$ but can be used more generally as long as anisotropy effects remain negligible.

For an ionic state, within a multiplet of total angular moment $J$, the MFF can be written, using the notations of Ref. [2],

$$\mu f(Q) = (\mu_L + \mu_S)\langle j_0(Q)\rangle + \mu_L \langle j_2(Q)\rangle, \qquad (1)$$

where $\mu = g_J \mu_B J$ is the total magnetic moment, $\mu_L = (2 - g_J)J$ and $\mu_S = 2(g_J - 1)J$ are its orbital and spin components, and the $\langle j_n(Q)\rangle$ are integrals of the product of the spherical Bessel function of order $n$ and the radial distribution of the electron density, derived from a Dirac-Fock relativistic calculation. Eqn. (1) can be rewritten in the familiar form

$$f(Q) = \langle j_0(Q) \rangle + C_2 \langle j_2(Q) \rangle, \qquad (2)$$

with

$$C_2 = \mu_L / \mu = (2 - g_J) / g_J, \qquad (3)$$

From Eqn. (1), it follows that the contribution of $\langle j_0(Q) \rangle$ to the MFF is associated with both the spin and orbital moments, whereas that of $\langle j_2(Q) \rangle$ results from the orbital moment alone.

The above dipole approximation has been shown to yield reliable results for the lanthanides [19]. This was confirmed, in particular, by numerous polarized neutron diffraction measurements, in which the magnetic moments are aligned by an applied magnetic field. Most $RE^{3+}$ ions exhibit a similar monotonic decrease of the static MFF $f(Q)$, because the values of $C_2$ are positive and generally comprised between 0 (for spin-only $Gd^{3+}$) and 2. One notable exception is, precisely, the case of $Sm^{3+}$, for which the calculated value of $C_2$ is as large as 5.42 [20] ($C_2 = 6$ in the dipole approximation). The physical origin is the pronounced cancellation taking place between the orbital ($\mu_L = 12/7\ J$) and spin ($\mu_S = -10/7\ J$) magnetic moment components in the $J = 5/2$ ground state multiplet. This results, as mentioned above, in a nonmonotonic function $f(Q)$ going through a maximum around $Q \sim 5$ Å$^{-1}$.

At this point, it is important to note that the general form of the magnetic spectral response [21], expressed in terms of the dynamic susceptibility, can actually be divided into Curie-type terms, associated with elastic and quasielastic scattering, and Van-Vleck terms, associated with CEF (*intra*multiplet, $\Delta J = 0$) and spin-orbit (*inter*multiplet, limited to $\Delta J = \pm 1$ in the dipole approximation) transitions. The scattering function is given as usual by

$$S(\vec{Q}, \hbar\omega, T) = (1/2\pi)(g_N r_e / \mu_B)^2 \chi''(\vec{Q}, \hbar\omega, T) / (1 - e^{-\beta\hbar\omega}), \qquad (4)$$

in which $\chi''(\vec{Q}, \hbar\omega, T)$ is the imaginary part of the dynamic susceptibility. Using the Kramers-Kronig relation, the latter quantity can be written [21]

$$\chi''(\vec{Q}, \omega, T) = \pi\hbar\omega \left[ \sum_m f_m^2(\vec{Q}, T) P_{mm}(Q, \hbar\omega, T) \chi_C^m(T) + \frac{1}{2} \sum_{m \neq n} f_{mn}^2(\vec{Q}, T) P_{nm}(Q, \hbar\omega - \Delta_{nm}, T) \chi_{VV}^{mn}(T) \right], \quad (5)$$

where the $P_{nm}(Q, \hbar\omega, T)$ are spectral functions (*e.g.* Lorentzian, Gaussian, delta, etc.), whose integral over energies must be equal to 1, and $f_{nm}(\vec{Q}, T)$ are magnetic form factors taking into account magnetic correlation effects in general case. $\chi_C^m(T)$ and $\chi_{VV}^{nm}(T)$ represent Curie (within energy level $m$) and Van Vleck (between levels $m$ and $n$) terms in the bulk static susceptibility, given by

$$\chi_C^m(T) = (g_J \mu_B^2 / k_B T) |\langle m | \hat{J}_z | m \rangle|^2 \exp(-\beta E_m) / Z \qquad (6)$$

and

$$\chi_{VV}^{nm}(T) = 2 g_J \mu_B^2 \frac{|\langle n | \hat{J}_z | m \rangle|^2}{\Delta_{nm}} \exp(-\beta E_m) / Z. \qquad (7)$$

Depending on the particular conditions of an experiment, a different functional dependence of the form factor $f(Q)$ may be involved. We have shown above the expressions valid for the static case (Bragg component). In the case of INS, a distinction must be made between QE contributions, corresponding to Curie terms, and excitations at finite energies corresponding to Van-Vleck terms in Eqn. (5). This can be a source of considerable dependence of the MFF on external parameters, in particular temperature as it was demonstrated in Refs. [8,22,23].

In the case of IV compounds, CEF transitions are usually wiped out by the strong valence fluctuations and therefore unobservable in INS experiments. They will not be considered hereafter. Spin-orbit intermultiplet ($J\leftrightarrow J\pm 1$) transitions, on the other hand, are related to a Van-Vleck-type magnetic susceptibility. The associated form factor has been shown [19,17] to be equal to $\langle j_0(Q)\rangle - \langle j_2(Q)\rangle$, i.e. corresponds to $C_2 = -1$ in Eqn. (2). According to realistic calculations of the radial electron density distribution, the values of $\langle j_0(Q)\rangle$ and $\langle j_2(Q)\rangle$ for neighboring 4f configurations, such as $f^5$ ($Sm^{3+}$) and $f^6$ ($Sm^{2+}$) [7], differ only slightly. Therefore, identical values of $C_2$ imply very similar $Q$ dependences of the MFF, as found experimentally for the spin-orbit transitions measured in $SmB_6$ [9,10,13], as well as in Sm(Y)S [14]. A similar character of the $Q$ dependence has been observed [24] for the $J=0 \rightarrow J=1$ excitation of integral valence $Sm^{2+}$ in pure semiconducting SmS. The peaks observed in the above experiments are associated with dipole transitions from the ground state to an excited state, which correspond to a pure Van-Vleck-type contribution to the dynamical susceptibility.

The QE signal, on the other hand, can be expected to provide information about the properties of the wave function of thermally populated states. In contrast to diffraction studies of the *induced* MFF, which have been mentioned above, it probes the undisturbed wave function without the application of an external magnetic field.

**2.2 MFF associated with QE neutron scattering in $SmB_6$**

Let us now consider the Curie-type contribution responsible for the quasielastic neutron scattering in $SmB_6$ at $T = 100$ K. We use the excitonic model of Ref. [16], which elaborates upon ideas first introduced by Stevens [25]. This approach is based on a trial wave function for the IV state of Sm, which qualitatively accounts for the main features of both the magnetic spectral response and lattice dynamics in $SmB_6$ and SmS.

The central point is the formation of an IV ground state wave function $\tilde{\Psi}_g$ consisting of an antisymmetrized product of quantum mechanically mixed states formed at each Sm site $m$,

$$\psi_{m,g} = \cos\theta \left| f_m^6, {}^7F_0 \right\rangle + \sin\theta \left| f_m^5 B_m^{(f)}, {}^7F_0 \right\rangle, \qquad (8)$$

where $B_m^{(f)}$ describes a loosely bound ("exciton") state of the electron located in the vicinity of the Sm ion. $\tilde{\Psi}_g$ thus represents a superposition of two initial states, $f_m^6$ and $f_m^5 B_m^{(f)}$, associated with the competing configurations $Sm^{2+}$ and $Sm^{3+}$. In Ref. [16], it was argued that the existence of this spatially extended component in the wave function was indeed responsible for the steeper $Q$ dependence observed in the intensity decrease of the 15-meV excitation peak at low temperature.

The integral-valence states $\left| f_m^6, {}^7F_0 \right\rangle$ and $\left| f_m^5, {}^6H_{5/2} \right\rangle$ themselves, which are no longer eigenstates in the low-temperature IV regime, are nonetheless clearly visible in the form of broad inelastic peaks in the experimental spectral response [9,10,13,14].

The suppression of the spin-gap when the system is heated to $T \geq 80$ K leads to a new regime in which the exciton peak is suppressed and a QE signal is recovered. This suggests that the magnetic response now arises from the fluctuations between the ${}^7F_J$ ($Sm^{2+}$) and ${}^6H_J$ ($Sm^{3+}$) parent configurations. Accordingly, the $Q$ dependence can be expected to take the form

$$f_{QE}^2(Q,T) = \sum_i v\, \sigma_{2,i}\, \rho_{2,i}(T,J) f_{2,i}^2(Q,T) + \sum_i (1-v)\, \sigma_{3,i}\, \rho_{3,i}(T,J) f_{3,i}^2(Q,T), \quad (9)$$

where $v = \cos^2\theta$ is the effective $Sm^{2+}$ fraction related to Eqn. (8), $\sigma_{2,i}$ ($\sigma_{3,i}$) are the neutron magnetic scattering cross sections for the different spin-orbit $J$ multiplets of $Sm^{2+}$ ($Sm^{3+}$), $f_{2,i}$ ($f_{3,i}$) the corresponding MFF calculated according to Eqns. (2) and (3), and $\rho_{2,i}$ ($\rho_{3,i}$) their Boltzmann population factors.

The above expression is a straightforward phenomenological extension of the original model, obtained by assuming a weighted contribution from each thermally populated multiplet, to describe the QE (Curie) contribution to the dynamic magnetic susceptibility in the IV regime. The MFF for the mixed state of Sm can be calculated as a function of the temperature and the average valence and compared to the experimental QE response presented in the next section.

### 3. Experimental results and discussion

Figure 1 presents a comparison of the $Q$ dependences of the different components of the magnetic response: *i)* the field-induced Bragg elastic signal [6] from the polarized neutrons and *ii)* the 14 meV exciton mode at $T = 2$ K [9 - 11, 13], obtained in previous neutron scattering experiments, together with *iii)* the QE intensity measured at $T = 100$ K on a $^{154}Sm^{11}B_6$ single crystal. The latter data were collected using the three-axis spectrometer 2T at LLB-Orphée (Saclay), operated at fixed final energy $E_f= 14.7$ meV. The solid and dashed lines correspond to the results of calculations [7,17,22] for the squared static MFF $f^2(Q)$ at $T= 0$ for ionic $Sm^{2+}$ and $Sm^{3+}$. The curve for $Sm^{2+}$ practically coincides with the data points corresponding to the experimental [6] induced MFF. We recall that nearly the same curve is expected for the $Q$ dependences of the $Sm^{2+}$ and $Sm^{3+}$ intermultiplet transitions (Van-Vleck terms) according to Eqn. (2) taking $C_2=-1$ [17], which was confirmed by time-of-flight data obtained on $^{154}Sm^{11}B_6$ powder [10,13] (not reproduced in Fig. 1 for clarity).

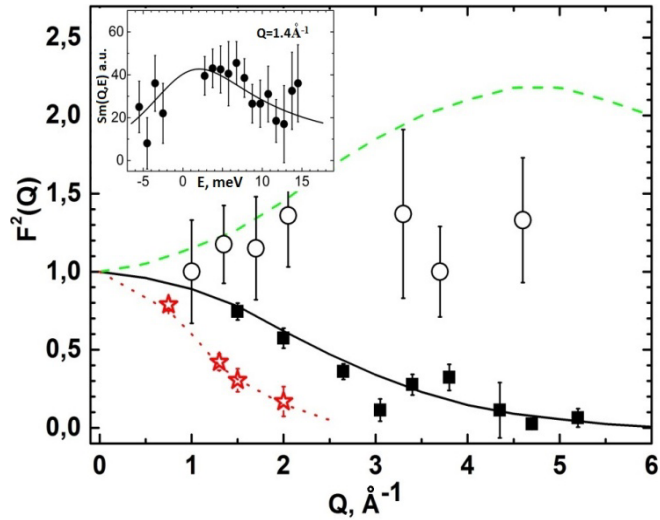

**Fig. 1.** *(Markers) Experimental Q dependences of different components of the magnetic response in IV $SmB_6$. Squares: polarized neutron diffraction in an applied magnetic field [6]; stars: resonance mode (14-meV excitation) at T=2 K [11]; circles: normalized quasielastic scattering intensity at T = 100K [this work]. Lines: calculated Q dependences of the static $Sm^{2+}$(solid) and $Sm^{3+}$ (dashed) MFFs [Eqns.(2) and (3)]. The dotted line is a guide to the eye through the star markers. (Inset) QE magnetic signal obtained as the difference between the neutron spectra S(Q,E) measured at T = 100K and 2K, for Q=1.4 Å$^{-1}$ at constant $E_f$=14.7meV (area of the elastic scattering near zero energy transfer is excluded). The line is a fit to the data using a Lorentzian spectral function with a line width $\Gamma_{QE}/2$=8(2) meV (half width at half maximum).*

The data in Fig.1 show a striking difference between the experimental $Q$ dependences of the different magnetic components. It is interesting to note that none of them exhibits the peculiar $Q$ dependence characteristic for the $J=5/2$ multiplet of $Sm^{3+}$.

The MFF derived from polarized neutron diffraction at $T = 7$ K in an applied magnetic field $H$ of 5 T (squares) is associated with a quite low magnetic moment of $\approx 0.02\mu_B$, as compared to the nominal magnetic moment of the ground state multiplet $J = 5/2$ in ionic $Sm^{3+}$ (about 0.8 $\mu_B$ ignoring possible CEF effects). Such a small value can be explained only by assuming that Sm has a singlet ground state, separated by a gap of the order of 20 meV [9-12], from a manifold of magnetic states. In this view, no contribution from any particular ionic state of $Sm^{3+}$ ($C_2 = 5.42$) is expected to exist in the induced MFF at low-temperature.

Another remarkable point is the steep intensity decrease of the exciton peak, which exists up to about 50K [11, 12]. This unconventional behavior was interpreted in as being due to an excitation from the IV Sm ground-state wave function $\tilde{\Psi}g$ (see above), and likely results from a spin-orbit type excitation involving the extended component [$B_m^{(f)}$ in Eqn. (8)] of the electron density [16].

From the above, it follows that the appearance of a QE signal in the excitation spectrum for $T \geq 80$K may provide the only experimental window for observing a Curie-type $Sm^{3+}$ component in the neutron spectra. Indeed, the experimental $Q$ dependence of the QE signal intensity differs qualitatively from all others, as can be seen in Fig. 1 (open circles, corresponding to spectra measured at $T = 100$ K), despite a relatively large scatter due to the low intensity and substantial broadening of the QE signal.

This experimental dependence can be compared to that predicted by Eqn. (10). Let us first note that, in using the dipole approximation [Eqn. (2)] for calculating $f_{2,i}$ and $f_{3,i}$, we ignore any intrinsic temperature dependence of those quantities (e.g. due to CEF effects). However, in view of the considerable uncertainty regarding the factors influencing, for instance, $f_{3,J=5/2}(Q)$ as a function of temperature [7,22], we believe that this simplification is not critical, though it could result in some overestimation of the contribution from the $J=5/2$ configuration in our calculation.

The question then arises of which value should be used for the calculation of Boltzmann population factors in Eqn. (10). One common phenomenological approach to describing IV systems consists in representing the effect of spin fluctuations (suppression of long-range magnetic order, non-diverging magnetic susceptibility for $T\rightarrow 0$, etc.) by an effective temperature $T_{eff}$, associated with the characteristic energy $E_{sf}$ of quantum spin relaxation in the 4$f$ electron system. $T_{eff}$ can be defined experimentally from the line width of the corresponding spectral component in the neutron spectra as $k_B T_{eff} = \Gamma/2$. In IV compounds, it may considerably differ from the actual (thermodynamic) temperature. For SmB$_6$, the characteristic spin fluctuation energy estimated from the line width of either the QE signal ($\Gamma_{QE}/2k_B \approx 100$ K, see inset in Fig. 1) or the intermultiplet excitation peaks ($\Gamma_{inter}/2k_B \approx 200$ K at $T = 10$ K [10,13]), is significantly higher than the thermal relaxation energy at the measuring temperatures. This is in contrast to "normal" RE systems, in which thermal induced spin fluctuations are typically one to two orders of magnitude lower than $k_B T$.

**Table 1.** *Parameters used for calculating the $Q$ dependence of the QE intensity according to Eqn. (9)*

|  | Fraction | $J$ | $\sigma_m$ (barn) | $C_2$ |
|---|---|---|---|---|
| Sm$^{2+}$ | 0.5 | 1 | 2.7 | 0.33 |
| Sm$^{3+}$ | 0.5 | 5/2 | 0.4 | 6.00 |
|  |  | 7/2 | 6.2 | 1.42 |

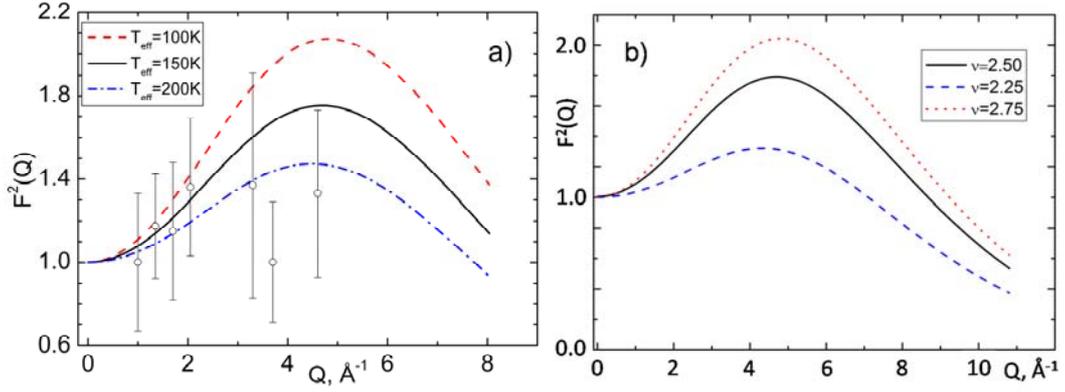

**Fig. 2.** *(a) Comparison of the experimental data from Fig. 1 for $SmB_6$ with the results of MFF calculations using Eqn. (10) (lines) for $T_{eff}$ = 100, 150, and 200 K. (b) Influence of the average Sm valence on the squared form factor $f_{QE}^2(Q)$ calculated at fixed $T_{eff}$ = 150 K.*

After considering all experimental information available from various INS studies, we chose $T_{eff} \approx 150$ K as the most realistic estimate. For this effective temperature, the $Q$ dependence of the QE intensity obtained from Eqn. (10) is mainly defined by the properties of the $J=1$ ($Sm^{2+}$) and $J=5/2, 7/2$ ($Sm^{3+}$) multiplets, listed in Table 1. The corresponding curve (solid line), as well as those associated with the lower and upper limits, $T_{eff} \approx 100$ and 200 K, from the QE and inter-multiplet components, is drawn in Fig. 2(a). One sees that calculations using $T_{eff}$ in this range of values agree reasonably well with experimental data, while emphasizing the marked difference between the MFF for the QE signal and the other contributions to the spectra. The present data thus provide the first experimental observation of a well-identified $Sm^{3+}$-type contribution to the $Q$ dependence of the magnetic response in IV $SmB_6$.

Eqn. (10) can further be used to predict how the $f_{QE}(Q)$ depends on the Sm average valence. For $SmB_6$ a valence of 2.5 ($v = 0.5$) was assumed in the above calculation, but this value is known to decrease when Sm is substituted by La, and to increase when it is replaced by Ba or Ca [10]. In Refs. [10,13], these valence changes were shown to affect the magnetic excitation spectra, in particular the MFF of the exciton mode. Figure 2(b) shows the calculated $Q$ dependence of the QE signal intensity for three values of the average Sm valence according to Eqn. (10). Unsurprisingly, the maximum in $f_{QE}^2(Q)$ becomes more pronounced with increasing trivalent character. No experimental data are currently available to check this prediction but measurements are in principle possible. In reality, due to variation of $v$ a change in the value of $T_{eff}$ may also occur[1], which cannot be reliably estimated at present time.

### 4. Conclusion

The quasielastic contribution to the magnetic spectral response of intermediate valence $SmB_6$ has been studied by inelastic neutron scattering on a $^{154}Sm^{11}B_6$ single crystal. The character of its observed $Q$ dependence differs qualitatively from that reported previously for either the field-induced (static) Bragg component, derived from polarized-neutron flipping ratios, or other inelastic contributions (intermultiplet transitions, "exciton" mode). In particular, it appears that no sizable decrease of the quasielastic signal with increasing $Q$ takes place from $1\text{Å}^{-1}$ to $5\text{Å}^{-1}$, in contrast to the typical behavior observed in most rare-earth ions, which points to a role of the lower $J=5/2$ multiplet of $Sm^{3+}$.

---

[1] Connection between the valence and $T_{eff}$ was discussed in experimental and theoretical studies of Ce compounds. But for systems having *f*-electrons in both configurations (Sm, Eu) this is not really established and analyzed yet, as far as we know, and we both parameters here are treated as phenomenological, in analogy with the Sales-Wohlleben model [26]

According to existing models for the IV state of Sm [16,25,27], likely applicable to SmB$_6$, this apparent contradiction can be solved by noting that both the field-induced static form factor and the excitations to higher *J* multiplets are related to *Van-Vleck*-type susceptibility terms out of the IV singlet ground state, whereas quasielastic scattering corresponds to Curie terms, which can occur only within degenerate energy levels (ground state or thermally populated excited states). Such states can be provided by the lower multiplets of the quantum-mechanically mixed 4*f*$^5$ (Sm$^{3+}$) and 4*f*$^6$ (Sm$^{2+}$) configurations at intermediate temperatures (*T* = 100 K in the present experiments). A calculation based on a simple phenomenological model qualitatively reproduces the experimental *Q* dependence of the quasielastic scattering intensity, and predicts a gradual enhancement of the maximum in $f_{QE}^2(Q)$ in SmB$_6$ solid solutions as the Sm valence gets closer to 3+.

The authors are grateful to A.S. Ivanov, K.S. Nemkovskii, and L.A. Maksimov for discussions and useful comments. PAA wishes to acknowledge the experimental support and hospitality from LLB-Orphée. The work is partly supported by the RFBR grant 14-02-00272-a.